# Low amplitude part of the electron detector response and its significance in neutron angular correlation measurements.

L.Goldin  and  B.Yerozolimsky
Harvard University, Cambridge, MA 02138

Abstract

Results of a  study of the possibilities to reduce the low amplitude "tail" of the response characteristic of various electron detectors are presented. The main reason of such attribute of all detectors used in electron spectrometry is the energy of several hundreds KeV taken by the backscattered electrons . A simple method of rejecting the events when the electrons are backscattered  - the use of a second - "veto"- detector in anticoincidences with the main detector was investigated.  The low amplitude "tail" in the response curve could be  reduced  by a factor of 3 – 4 . The  remaining effect  - about 1 % of  the integral has yet no explanation.  Additional  experiments showed that only ~ 0.2-0.3 % can be related to bremsstrahlung.
The significance of this effect in the study of angular correlations in neutron beta decay was analyzed too with the help of a simplified computer model . As a result, we propose a method of calculating appropriate corrections which promise to reduce the systematic uncertainty in the measurement of the "a" correlation coefficient which  going to be carried out in the near future.

Spectrometric properties of various electron detectors have been investigated for many years, and it is well known that all of them have in their response curves some low amplitude region which  stretches  between the Gaussian peak and zero ( for instance [1 ] ).  The  integral of this number - low amplitude "tail" - depends upon   the detector material,  and in case of a lightest one - plastic scintillator -  it amounts about 4 % relative to the full area of the response curve.   The stretched out shape of the response curve in the low amplitude region is caused mainly by the backscattering of electrons from the detector leading to incomplete transfer of the particle energy to the detector.  In many experiments devoted to the problems of electron spectrometry this effect  plays a relative negligible role , particularly if compared with the importance of the main characteristics of a detector used in spectrometry – the energy resolution defined by the width of the Gaussian part of the response .

There are nevertheless some cases when this effect becomes very important and defines the systematic uncertainty which can be achieved in the experiment. Just such kind of experiments is the one being prepared nowadays devoted to a precise measurement of the constant "a" defining the angular correlation between the electron and antineutrino in free neutron beta decay. The study of this elementary decay process remains to be till now extremely important because it gives a valuable data for approving the predictions of the Week Interaction theory.

The main idea of this new experiment is based on the separation of two groups of events in the recoil proton time-of-flight spectrum. which correspond to opposite antineutrino escape directions The consideration of the momentum diagram of the decay products – electron, antineutrino and recoil proton –shows that the conditions of the proposed experiment provide a complete separation of these two kinds of decay events only if the electron energy spectrum is limited by a some maximum value [2 ], [3 ]. But it is evident, that the presence of a low amplitude "tail" in the detector response characteristic leads to admixing the electrons with energies exceeding this maximum limit with electrons below this limit. As a result, the separation of the groups in the proton time-of-flight-spectrum becomes incomplete which causes a systematic uncertainty in the value of the constant "a" obtained. Calculations based on a computer simulation of this experiment [3] showed that even in the case of a scintillation plastic as an electron detector this uncertainty in "a" becomes to be about 10 – 15 %.

This was the reason to undertake experimental studies to find a method of reducing of the "tail' part of the electron detector amplitude distribution , and these investigations started in the Hugh Energy Physics laboratory of Harvard University in 1995.

In so far as the main reason which causes this effect is connected with electron backscattering from the surface of the detector, we based our study on a simplest idea of rejecting events accompanied with scattered electrons with the help of an additional detector which had to pick up these scattered particles and activate the "veto" of an anticoincidence module. One of the last versions of the experimental set-up which we used is presented in Fig 1. The "veto" detector was a cylinder (6 ) made of a plastic scintilator BC-404 viewed by several photomultipliers ( there are two in Fig 1 ) and disposed before the main detector ( in Fig 1 it is a 30 mm diameter and 1.5 mm thick plate made of scintillation plastic BC-404 viewed by a PMT - 8850 ). This geometry provided a more than 95 % probability that the electron backscattered from the main detector will hit the inner cylindrical surface of the "veto" detector.

A conversion electron source Sn-113 ( E = 365 KeV ) was used in these experiments ( 1) , and the energy spectrum of electrons was additionally cleaned up with the help of a 180 degree magnetic separator (3 ) which filtered electrons with energies differing from 365 KeV and gammas from the source too. The beta- active isotope Sn-113 was collected in the sample of separated mono-isotope Sn-112 ( 95% enrichment) during the irradiation on the MIT reactor ( in the neutron flux ~ 5 $*10^{12}$ $1/cm^2 sec$ ). The activity of the source after irradiation was about 5* $10^5$ per sec.

The solenoid (5 ) shown in Fig 1 served for focusing the electrons on the active surface of the detector. As a result, the counting rate connected with the detecting of electrons of the source in our set up was about 30 counts/sec with a fresh irradiated source.

The half life period of Sn-113 is about 110 days, thus we had to reactivate the source from time to time at the MIT reactor.

The counting rate of the detector background was about 1 per sec, and in order to get the desired information about the real amplitude spectrum in the region of the "tail", where the counting rate is very low, we had to subtract the background spectrum, and having in mind all kind of instabilities of the detecting system ( high voltage of PMT , the multiplication of PMT itself, the gain of the electronic modules, the value of the magnetic field in the magnetic separator etc) we had to measure the spectrum related to the source electrons and the spectrum of the background as close to the simultaneous measurement as possible. In order to measure the background spectrum we simply switched out the magnetic field of the separator, and the intervals of time of the measurement " with the magnetic field " and "without the field" were made as short as ~ 2 minutes. The spectrums in both cases were collected in two separate files of the on-line computer and in separate integral counters. All procedures of switching on and switching out the current in the magnetic separator and switching the files and the integral counters were accomplished automatically with the help of a special computer program and control modules.

After finishing the data collection ( usually lasting 6 – 30 hours ) the obtained spectra were subtracted one from another so that the difference spectrum was only the spectrum of signals belonging to detected electrons of the source .

Several types of detectors have been investigated during the last years : plastic scintillators , liquid scintillators, monocrystaline – Stilbene and semiconductor Si.

Different variants of the geometry of the "veto" detector and different numbers of PMT connected with this detector were used in these experiments.

Besides, we were concerned about the real width of energy distribution of the electrons hitting the detector due to possible scattering from the edges of the diaphragms installed on the way of electrons which come from the Sn-113 source. In order to be sure that these effects do not affect the electron spectrum we changed the materials the diaphragms were made of, as well as their dimensions and positions in the chamber of magnetic separator . The results of these experiments lead to the conclusion that the diaphragms we used did not affect the low amplitude part of the electron spectrum on the level of several tenths of one percent.

The best shapes of the response curve of a beta detector which we could obtain are presented in Fig 2 and Fig 3. The main detector in these measurements was a plastic scintillator BC-404 ( diameter 30 mm and 1.5 mm thick) and the configuration of the "veto" detector was as shown in Fig 1 with two PMT viewing it.

Data presented in Fig 2 show the differential amplitude spectrum of the main detector with the "veto" detector switched in anti-coincidences with the main one . Fig 3 presents the "tail" parts of the spectra showed as integrals calculated from zero channel up to the channel on the abscissa relative to the whole integral in % .

These results confirm the basic idea of the important role which backscattering effect plays in creating the low-amplitude part of the response curve and show that this "tail" can be essentially reduced with the help of a "veto" detector. But at the same time it becomes reasonable to propose that there may be some other effects which are responsible for this part of the amplitude spectrum, because in spite of the fact that the "veto" system seems to be efficient enough to reject almost all events when electrons are backscattered, the amplitude distribution still has some residual "tail" of low amplitude signals on the level about 1 % ( if the integral is measured in the diapason $0 < E_e < 285 KeV$ ).

Just this fact made us to search for alternative types of detectors. The first proposition was to relate this additional mechanism of loosing some part of electron energy to some solid effects in the plastic scintillator ( like excitation) , and we decided to try a liquid scintillator as the main detector. It turned out however, to be a difficult task because the liquid scintillator we used can not exist if the pressure is less than ~50 torr ( it evaporates ) and the vacuum in the volume, where the electrons are moving, must be < $10^{-1}$ torr : besides, the film separating the liquid scintillator from the vacuum must be very thin, and any supporting grid is excluded due to scattering of electrons. Thus, we had to arrange the vacuum system with two separate volumes: one for the moving electrons ,where the vacuum was relatively high, and another for the liquid scintillator with ~ 100 torr pressure. The final variant of the construction of the detector with a liquid scintillator is shown in Fig 4. Unfortunately, the results of investigating this detector turned out to be disappointing : the quality of the response was practically the same as in the case of plastic scintillator.

The second hypothesis about the origin of the residual "tail" was the well known bremsstrahlung which must be emitted when the electrons are losing their velocity inside the detector[4 ]. In order to study this effect we used a Si semiconductor as the main detector and placed a CsI(Tl) scintillator close behind ( Fig 5 ).
 The PMT viewing this heavy scintillator which is sensitive to gamma rays was switched in coincidences with the main detector and ether the spectrum of Si detector or the spectrum of gammas detected by the CsI(Tl) was measured.
Fig 6 and Fig 7 present these both spectra, and they together with each other confirm the bremsstrahlung as the origin of coincidences . The measured counting rate of coincidences in these experiments was 0.25±0.05 % of the integral counting rate of the Si detector. Taking in account the effective solid angle of registration of gammas emitted from the Si-detector ( ~ 0.3 of $4\pi$ ) and the ratio between the probabilities of bremsstrahlung effect in Si and plastic one can estimate that the number of events when electrons lose more than ~50Kev due to the bremstrahlung in the plastic scintillator is less than ~ 0.3% of the integral counting rate of the detector.

Thus, the origin of more than 0.5 % in the "tail " of plastic scintillator response remains still unaccountable. Either it is an intrinsic property of the detectors investigated or it is connected with some defects of our experimental set-up which cause a real spread of the energy spectrum of electrons – this remains to be a question which must be investigated further.

Nevertheless, it seems to be worthwhile to try to find some method of treatment of experimental data derived in the proposed investigation of angular correlation in neutron decay that could give a possibility of taking in account the effect connected with the low amplitude "tail" in the response characteristic of beta detector if this "tail" will be the same as we achieved in our best results ( demonstrated in Fig 2, Fig 3).
Trying to solve this problem we used a simplified model of our neutron decay experiment ( point-like decay region and point-like beta detector) which could permit us to derive the time-of-flight spectra of the recoil protons without the Monte Carlo calculations. A much more simple and fast program was created . It is based on some analytical equations and comes to consecutive selecting of various combinations of several parameters characterizing the neutron decay process (the energy of the electron, the energy channel of the beta detector which is defined by its response, including the "tail" of low amplitudes and the angles of antineutrino escape direction ) . The calculations based on the use of this program takes several seconds, and thus, we could change the conditions of the experiment many times. We chose the energy interval of the beta detector, its energy resolution , the value of the "tail" in the electron detector response and even the shape of this "tail" ( dividing the whole energy diapason between zero and electron energy in five equal intervals and chousing the coefficients defining their relative parts in the "tail") and other parameters of the experiment too.
Using the results of these investigations we could propose a simple procedure of treating the experimental spectrums in order to calculate the desired angular correlation coefficient "a" . The analysis carried out with this simplified approach shows that if the shape of the electron detector response is like the one demonstrated in Fig 3, the correction in the calculated value of "a" which must take in account the "tail" of this response is about 5% and the systematic uncertainty of this correction might be as low as ±0.4%.
The detailed description of this method of treating the experimental data will be done in a separate paper.

# CONCLUSIONS

Our experiments showed that the method of suppressing the electron backscattering effect based on the use of an additional Veto –detector switched in anticoincidences with the main detector reduces the low amplitude "tail" in the response curve of electron detectors. The best results were achieved with the plastic scintillator: the "tail" was reduced by a factor 3 – 4 , and the residual integral of low amplitude signals ( ~ 1 % of the full integral ) is partly connected with the bremsstrahlung effect (~0.2 –0.3 % ). The nature of the remaining part of the "tail" is not understood yet and will be investigated later.

Computer simulation of the experiment devoted to the measurement of the angular correlation constant "a" in neutron decay was carried out. A method of taking in account the false asymmetry caused by the presence of the "tail" in the beta detector response we could not completely suppress was proposed, and it seems that the systematic uncertainty in "a" connected with this procedure will be on the admissible level.

Acknowledgments

The authors are very thankful to Prof. Richard Wilson for his valuable support of our experimental efforts during all these years.
A very important part of the work we are presenting in this article – creating of all computer programs and help in solving electronic problems has been done by Aleksey Yerozolimsky , and we are grateful for it.
We appreciate many discussions about the results we obtained with Dr Maynard S.Dewey and Prof. Fred E. Wietfeldt which helped us during these investigations.

FIGURE CAPTIONS

Fig 1  Experimental set-up at Harvard University
    1 – Sn-113 electron source E = 360 KeV ;  2 – Pb shield ;  3 – magnetic separator
    with thin copper diaphragms and magnetic field ~ 270 Gs  ; 4 – steel magnetic
    shield ; 5 – focusing solenoid ;  6 – "veto" detector – plastic scintillator with two
    PMT;  7 – main detector – plastic scintillator BC-404 1.5 mm thick,  30 mm diam.;
    8 – light guide  ( absent in the last version of the set-up) ;  9 – reflector – Al foil.

Fig 2  Differencial spectrum derived with the plastic scintillator BC – 404.
    Veto signals switched in anticoincidences.
    Symmetric distribution is derived by a simple mirror reflection of the right slope,
    thus,  the exceeding part of the experimental spectrum is just the "tail" of low
    amplitude signals .

Fig 3  Integral spectra of the "tail"  in % of  the whole integral.
    1 – without anticoincidences with the signal of the "veto" detector;
    2 – with the anticoincidencees

Fig 4  Construction of a detector with the liquid scintillatorm

Fig 5  Search for the bremsstrahlung gammas.

Fig 6  Spectrum of electrons in the Si detector ( coincidences with gammas ) .

Fig 7  Spectrum of gammas in the CsI (Tl) detector ( coincidences  with electrons).

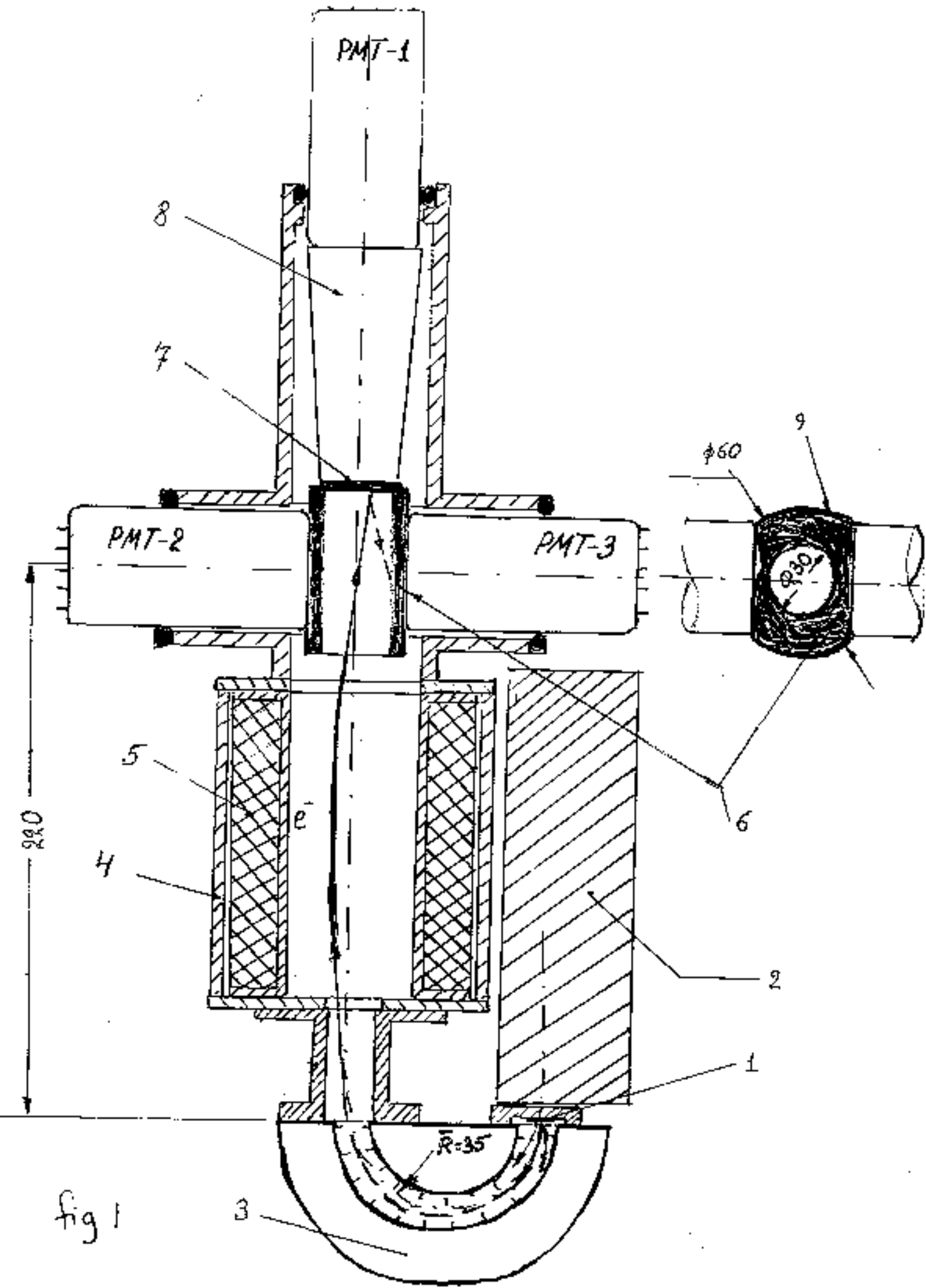

fig 1

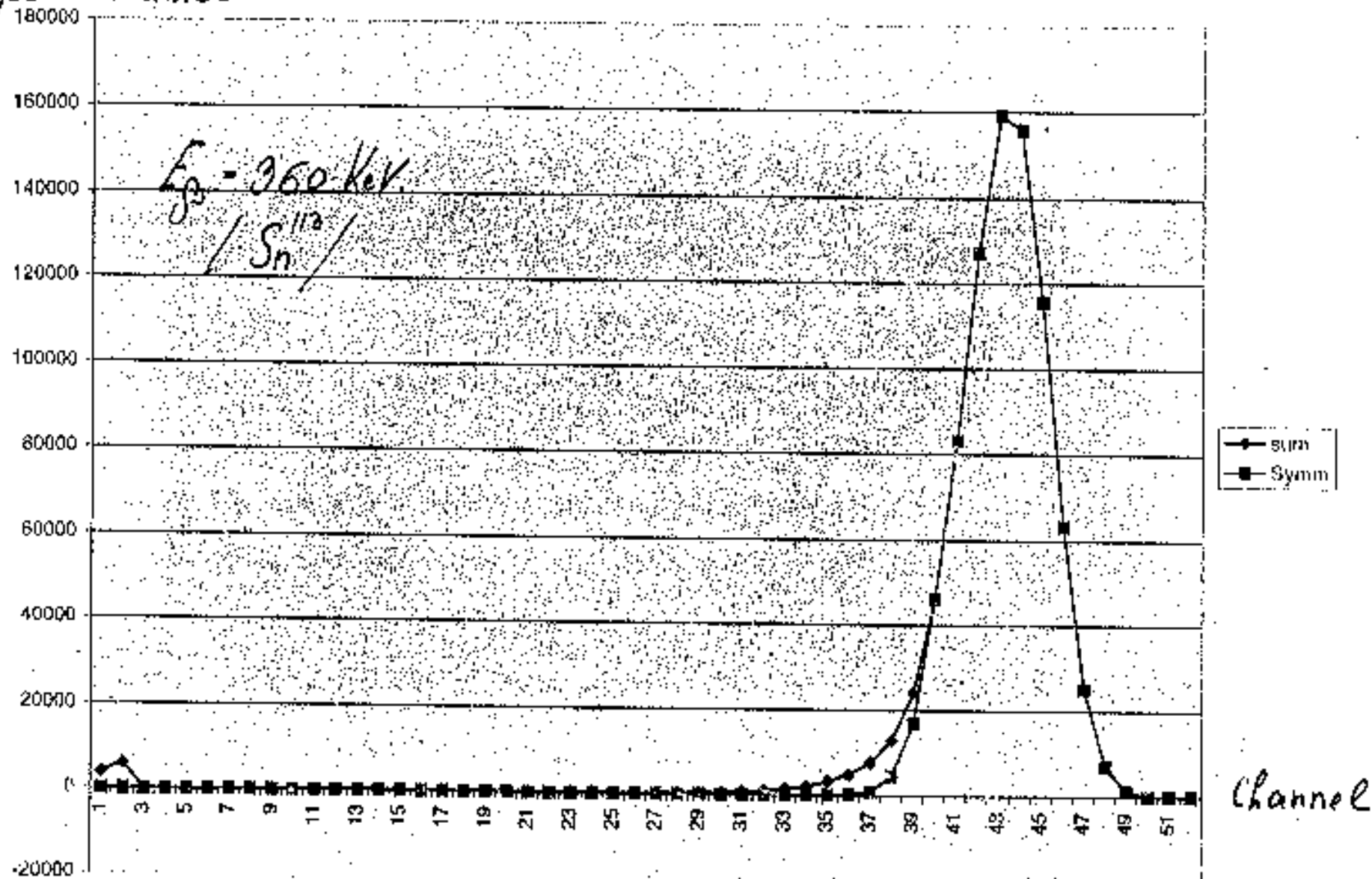

fig 2

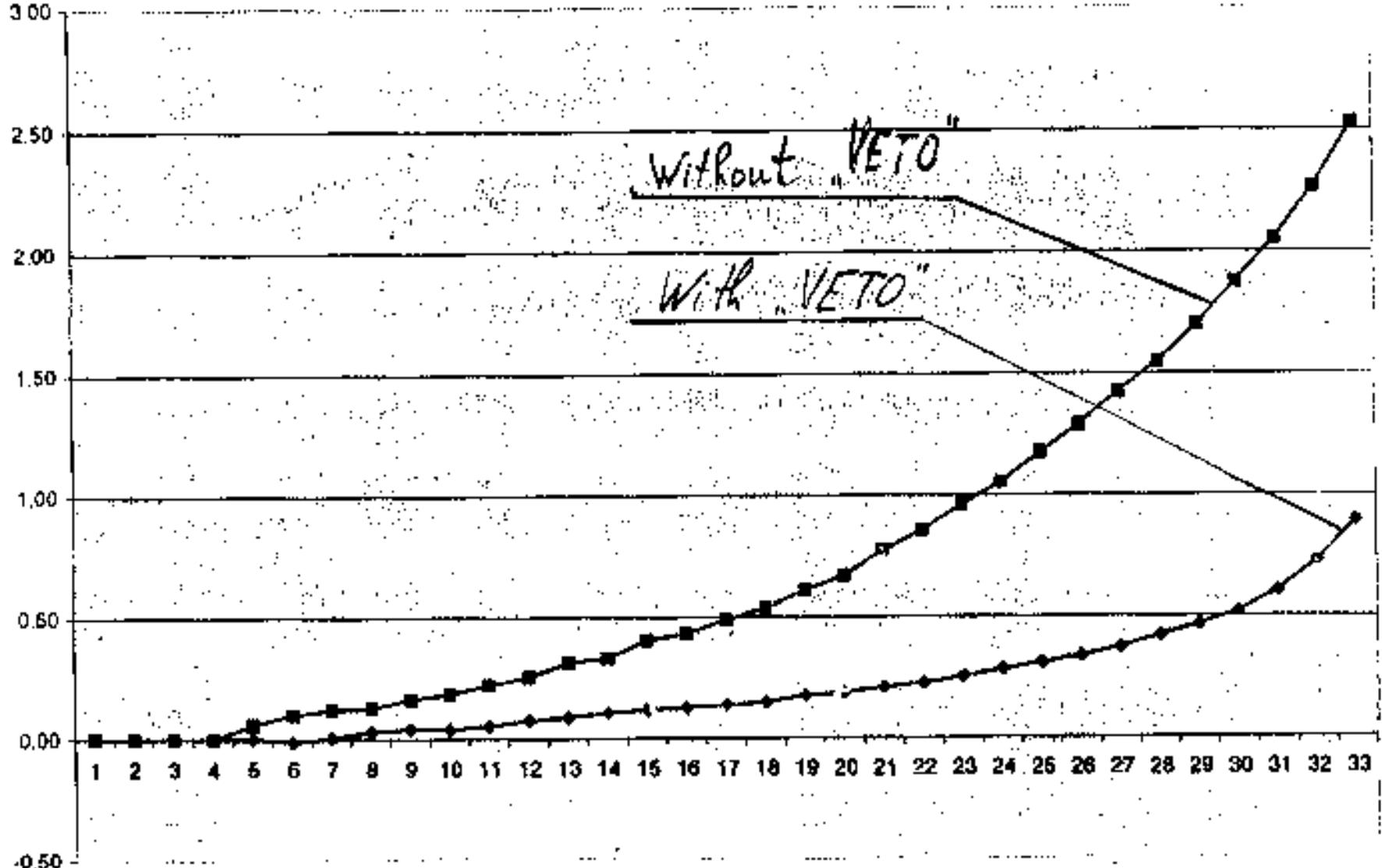

fig 3

Plastic Scintillator

Max in Ch. N-43

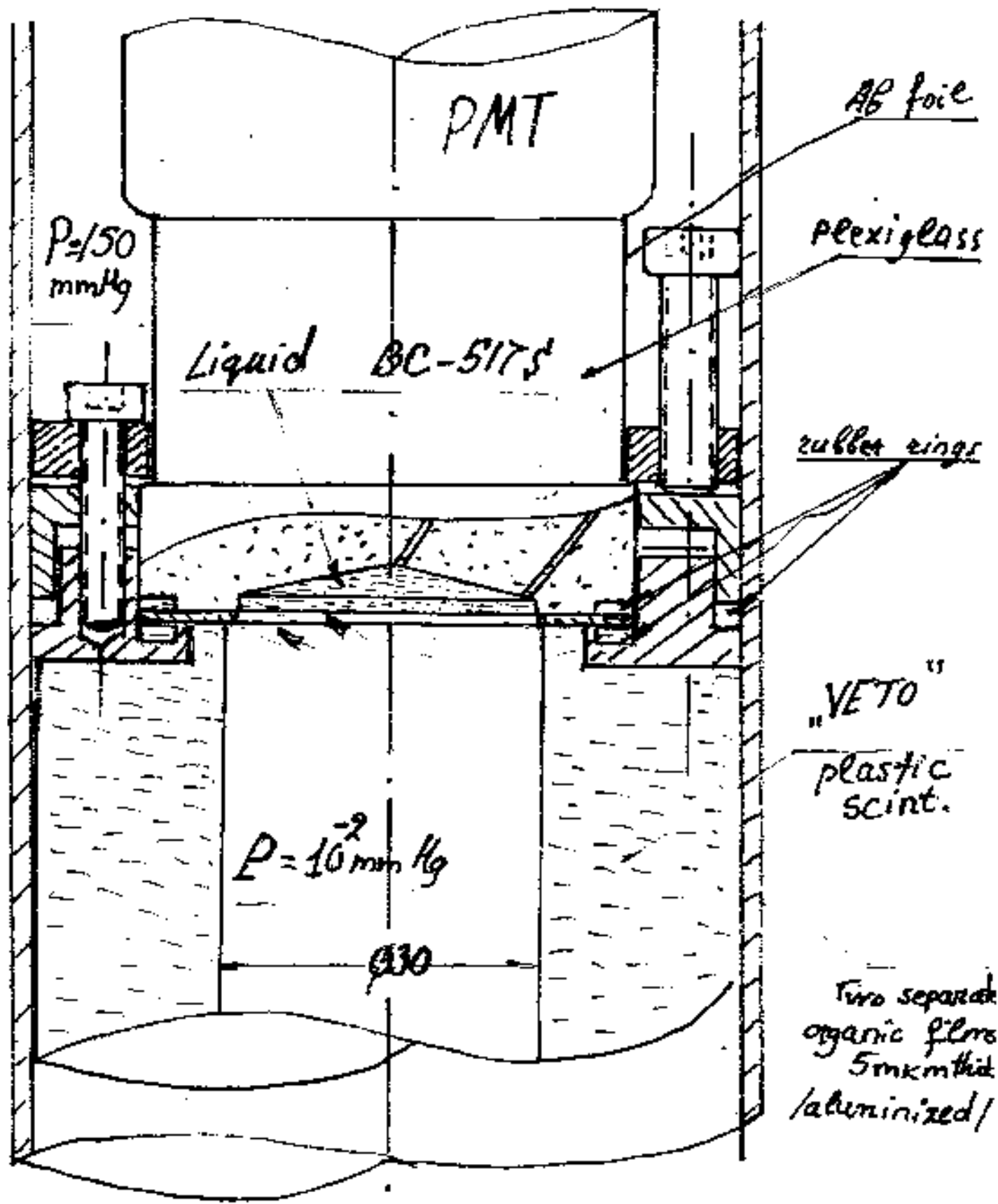

fig 4
Construction of the liquid scintillator  M 2:1

# Search for bremsstrahlung gammas

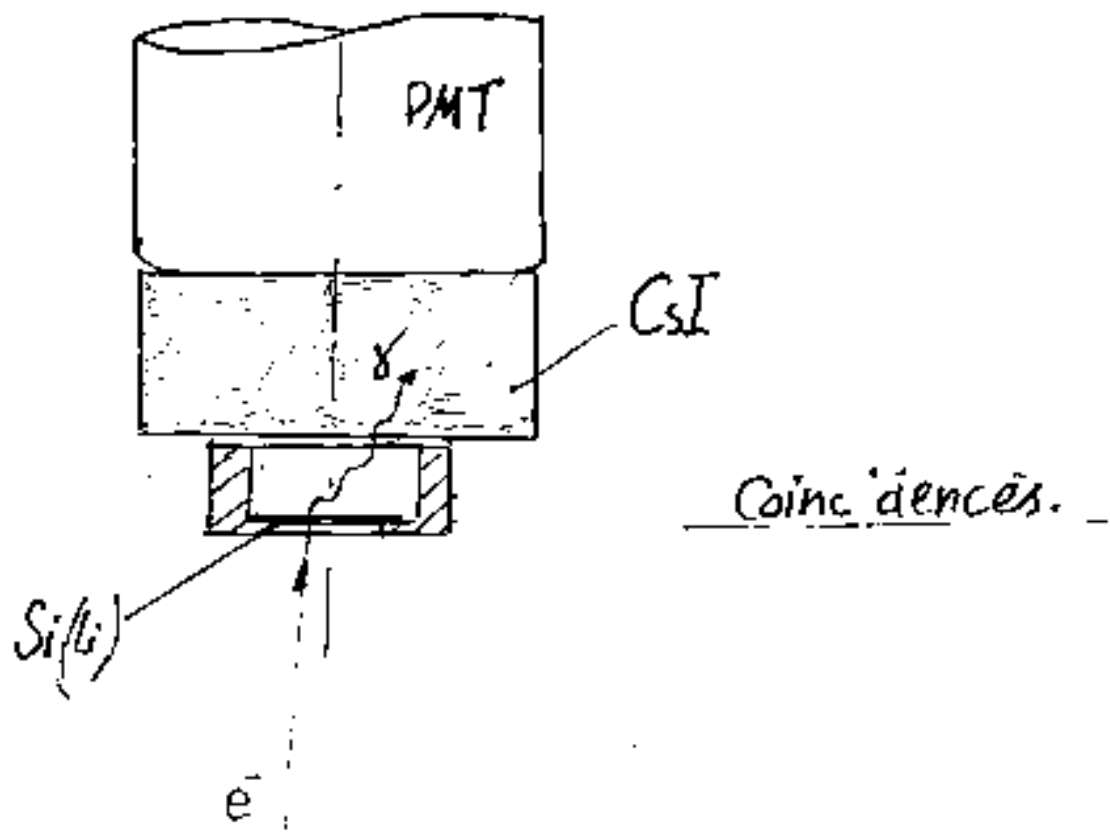

Coincidences.

fig 5

August 2002

# Si(Li) - SPECTRUM OF ELECTRONS IN COINCIDENCES WITH CsI DETECTOR OF OF GAMMAS

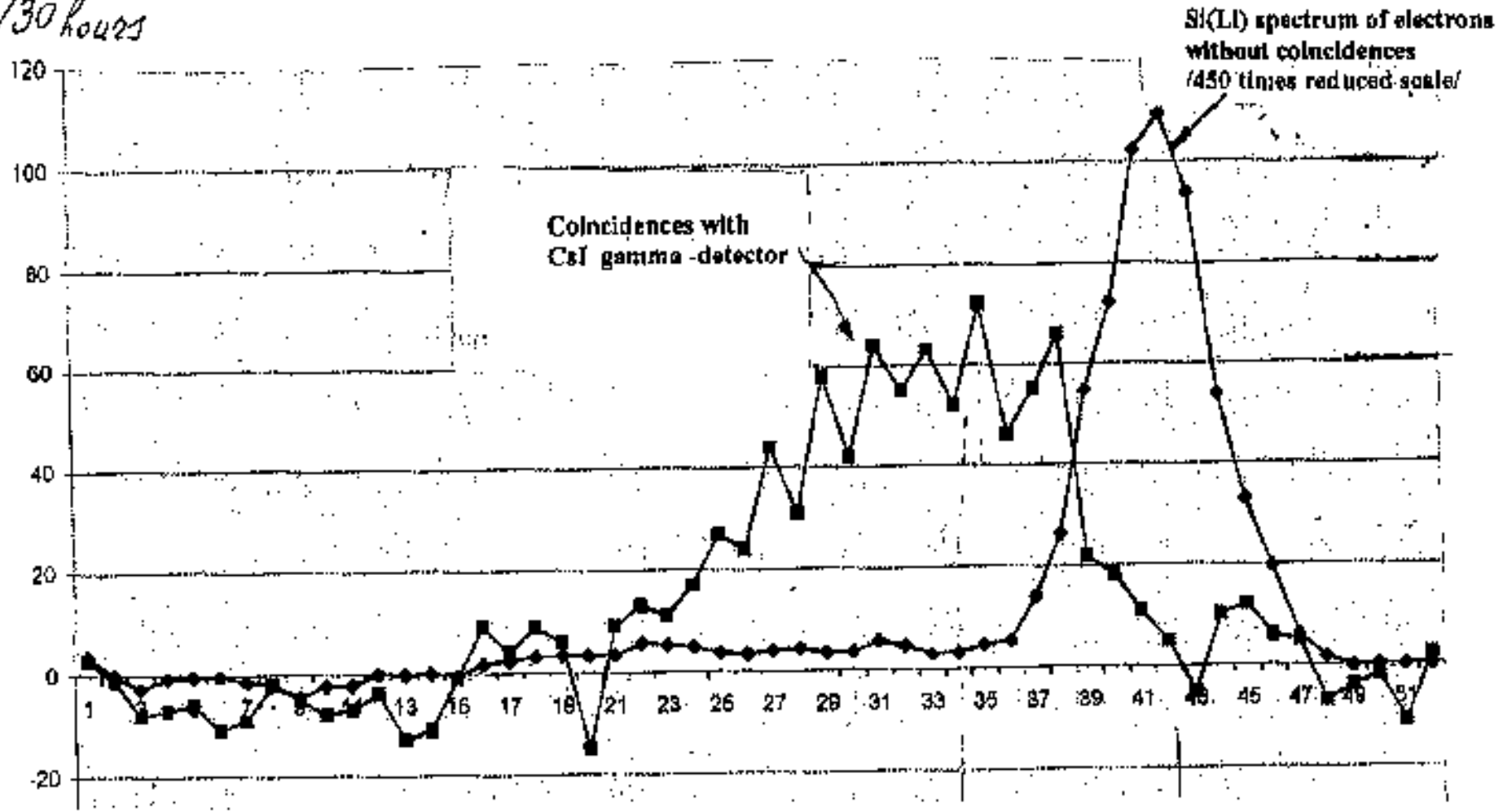

Fig. 6

August 2002

N/5hours

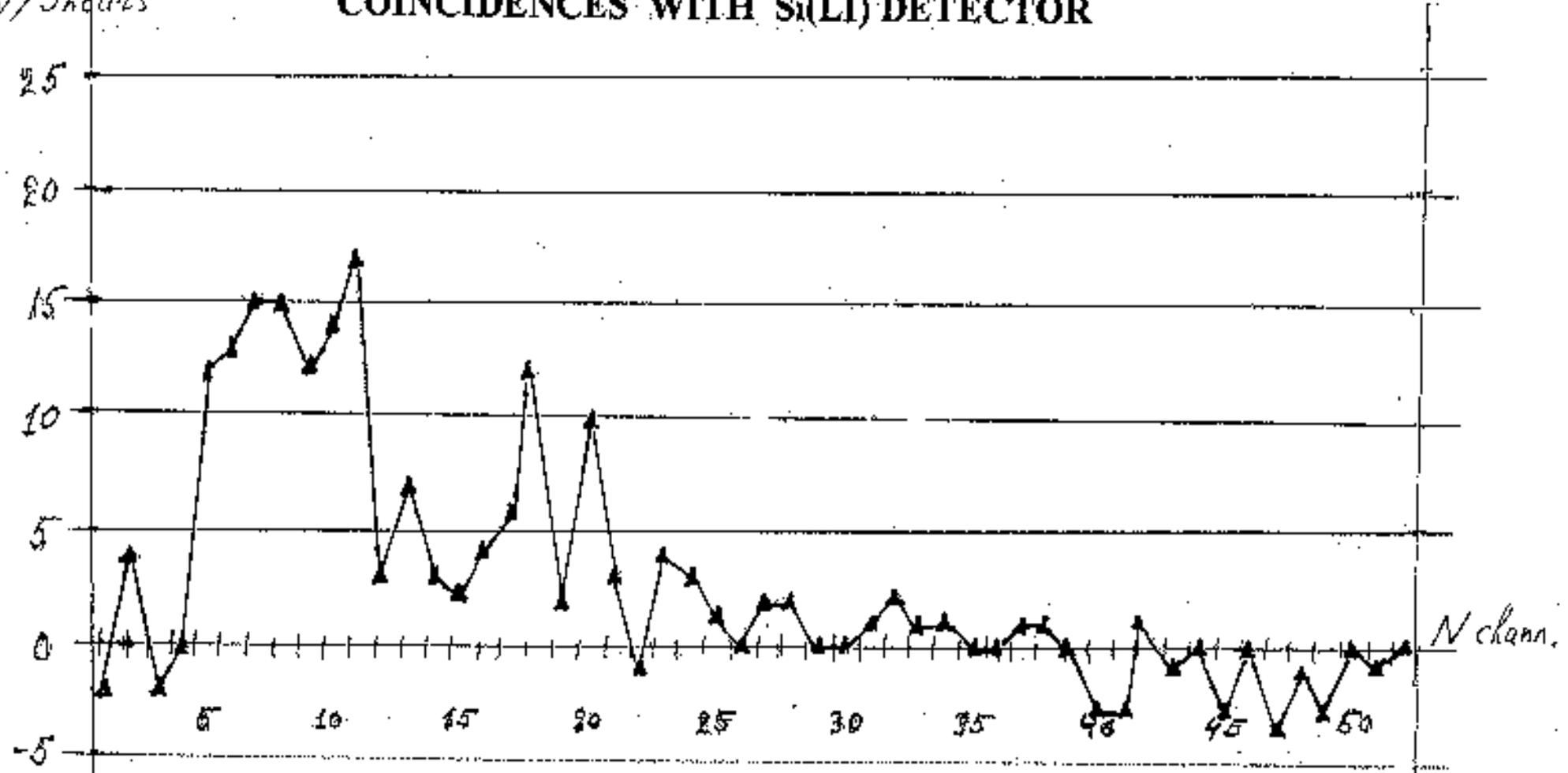

CsI SPECTRUM OF GAMMAS IN COINCIDENCES WITH Si(LI) DETECTOR

fig 7